\documentclass[12pt,preprint]{emulateapj}

\usepackage{amsmath}

\slugcomment{Accepted by the Astrophysical Journal}

\shorttitle{Temperature and abundance profiles in hot Galaxy Clusters}
\shortauthors{Baldi et al.}

\begin{document}

\title{A Chandra archival study of the temperature and metal abundance profiles
in hot Galaxy Clusters at $0.1\la z\la0.3$}


\author{A. Baldi}
\affil{Harvard-Smithsonian Center for Astrophysics}

\author{S. Ettori}
\affil{INAF - Osservatorio Astronomico di Bologna}

\author{P. Mazzotta\altaffilmark{1}}
\affil{Universit\`a di Roma "Tor Vergata", Dip. di Fisica}

\author{P. Tozzi\altaffilmark{2}}
\affil{INAF - Osservatorio Astronomico di Trieste}

\author{S. Borgani\altaffilmark{2,3}}
\affil{Dipartimento di Astronomia dell'Universit\`a di Trieste}


\altaffiltext{1}{Harvard-Smithsonian Center for Astrophysics}
\altaffiltext{2}{INFN - Sezione di Trieste}
\altaffiltext{3}{INAF - Osservatorio Astronomico di Trieste}

\begin{abstract}
We present the analysis of the temperature and metallicity profiles
of 12 galaxy clusters in the redshift range 0.1--0.3 selected from
the Chandra archive with at least $\sim20,000$ net ACIS counts and
$kT>6$ keV. We divide the sample between 7 Cooling-Core (CC) and 5 Non-Cooling-Core
(NCC) clusters according to their central cooling time. 
We find that single power-laws can describe properly both the 
temperature and metallicity profiles at radii larger than $0.1 r_{180}$ in 
both CC and NCC systems, showing the NCC objects steeper 
profiles outwards.
A significant deviation is only present in the inner $0.1 r_{180}$.
We perform a comparison of our sample with the De Grandi \& Molendi
BeppoSAX sample of local CC and NCC clusters, finding a complete agreement in the
CC cluster profile and a marginally higher value (at $\sim1\sigma$) in the
inner regions of the NCC clusters.
The slope of the power-law describing $kT(r)$ within $0.1 r_{180}$
correlates strongly with the ratio between the cooling time and the age
of the Universe at the cluster redshift, being the slope $>0$ and
$\tau_c/\tau_{age}\la0.6$ in CC systems.
\end{abstract}



\section{Introduction}
Clusters of galaxies represent unique signposts in the Universe, where
the physical properties of the cosmic diffuse baryons can be
studied in great details and used to trace the past history of cosmic
structure formation (e.g. Rosati et al. 2002; Voit 2005, for reviews).
As a result of adiabatic compression and shocks generated by
supersonic motion during shell crossing and virialization, a hot thin
gas permeating the cluster gravitational potential well is
formed. Typically this gas, which is enriched with metals ejected form
supernovae (SNe) explosions through subsequent episodes of star
formation (e.g. Matteucci \& Vettolani 1988; Renzini 1997), reaches
temperatures of several $10^7$ K and therefore emits mainly via
thermal bremsstrahlung in the X-rays. At such temperatures most of the
elements are either fully ionized or in a high ionization
state.\\
Particularly evident in X--ray spectra of galaxy clusters are the
strong transitions to the $n=1$ level (K--shell) of the H--like and
He--like ions of Iron in the energy range 6.7--6.9 keV.  Below 2 keV,
$n=2$ level (L--shell) transition of Iron and $\alpha$ elements can be
detected, especially in the low temperature region in the centers of
the so--called cool core clusters, that are characterized by a
strong peak in the surface brightness distribution, and therefore
short cooling times.
Spatially resolved CC clusters show a peak in the metal distribution
associated to the low temperature core region
(e.g. De Grandi \& Molendi 2001 and 2002, hereafter DM01 and DM02).
Although the amount of energy supplied to the
intra-cluster medium (ICM) by SNe explosions depends on several
factors (e.g. the physical condition of the ICM at the epoch of the
enrichment) and cannot be obtained directly from X--ray observations,
the radial distribution of metals, as well as their abundance as a
function of time, are crucial information to shed light on the cosmic
star formation history and to trace the effect of SN feedback on the ICM.\\
Several analyses have been presented in the literature with the aim to
study the radial distribution of metals in clusters of galaxies.
Finoguenov, David \& Ponman (2000) performed a spatially resolved
X-ray spectroscopic analysis of 11 relaxed clusters observed by ROSAT
and ASCA, deriving a radial distribution of single heavy elements such
as Fe, Si, Ne and S.  They found that the total Fe abundance decreases
significantly with radius in all clusters, while the Si, Ne and S
abundances are either flat or decrease less rapidly. DM01 derived
radial metallicity profiles (mainly driven by Fe) of 17 nearby
clusters observed by BeppoSAX. They found a strong enhancement in the
abundance in the central regions of the CC clusters. A flatter
metallicity profile was observed instead for the non-cool core 
clusters in their sample. Since all the NCC clusters show signs of
recent merger activity, they suggested that the merger events may have
redistributed efficiently the metal content of the intracluster
medium.  Irwin \& Bregman (2001) derived iron-abundance profiles for
12 clusters with $0.03\le z\le0.2$ observed by Beppo SAX. Although
they investigated the differences between CC and NCC clusters in a
less systematic way than DM01, they found a negative gradient in the
abundance profiles of all the CC clusters and to a lesser significance
also in the NCC clusters. Similarly to DM01, they found that CC
clusters have higher metallicity than NCC clusters at every radius. It
is worth to say that the aforementioned papers investigated the
metallicity trends only within $r_{500}$.
Spatially resolved measures of the metal abundance in galaxy clusters
were performed also with XMM-Newton. In particular Tamura et al. (2004)
analyzed a sample of 19 X-ray bright relaxed clusters, obtaining elemental
abundances of Fe, Si, S and O. They found that while the distribution of
Fe, Si and S is generally peaked toward the center, the O abundances are
uniform throughout the cluster, pointing out to a different origin among
these metals, most likely in SNe Ia and II.
More recently Vikhlinin et
al. (2005) have derived temperature and metallicity profiles for 11
low-redshift clusters observed by Chandra. The clusters in their
sample however are all CC clusters, presenting a very regular overall
X-ray morphology and showing only very weak signs of dynamical
activity. Although they have not analyzed the metallicity profiles
rescaled to the virial radius of the cluster (as they did for the
temperature profiles) a negative gradient of $Z$ is
present in all the objects in their sample.
However, almost all the spatially-resolved metallicity profiles are measured
in local clusters ($z<0.1$).

On the other hand, measurements of the metal content of the intracluster medium
at high$-z$ has been obtained with single emission-weighted estimates from Chandra
and XMM-Newton exposures of 56 clusters at $0.3\la z\la 1.3$ in Balestra et al. (2007).
They measured the Iron abundance within (0.15-0.3)$R_{vir}$ and found
a negative evolution of $Z_{Fe}$ with the redsfhift, with clusters at $z\ga 0.5$ having
a constant average Fe abundance of $\approx0.25Z_\odot$, while objects
in the redshift range $0.3\la z \la0.5$ show $Z_{Fe}$ significantly higher
($\approx0.4Z_\odot$). Such evolution is not driven entirely by the presence 
of the cool cores.
This result has been recently confirmed by Maughan et al. (2007).

In this paper, we present measurements of the radial temperature and
metallicity profiles of a sample of 12 clusters with temperatures larger than
6 keV observed with Chandra at intermediate redshift, $0.11\le z\le0.32$.
We take advantage of the ACIS superior spatial and spectral resolution to
investigate in a systematic fashion the differences that may exist
between CC and NCC clusters.
The spectroscopic measurements of the ICM temperature and metallicity
allow to characterize statistically the radial profiles
and to quantify their gradients in this unexplored
redshift region.

All the uncertainties are quoted at 1$\sigma$ (68\%) for one
interesting parameter. The abundance estimates are relative to the
compilation of cosmic values given in Anders \& Grevesse (1989)
(hereafter AG89), unless otherwise stated.  Indeed, these values for
the solar metallicities have more recently been superseeded by the new
values by Grevesse \& Sauval (1998) and Asplund et al. (2005)
(hereafter A05), who introduced a 0.676 and 0.60 times lower Iron
solar abundance, respectively (photospheric value), while the other
elements do not change significantly.  Our measures of metallicity are
expected to be driven mainly by Iron, however, for clarity, we also
performed the fits using solar abundances by A05.
\\
Throughout this paper we assume $H_0=100\:h$ km s$^{-1}$ Mpc$^{-1}$,
$h=0.7$, $\Omega_m=0.3$ and $\Omega_\Lambda=0.7$.

\section{Sample Definition and Data Analysis} \label{dataprep} From
Chandra archival data we select a sample of twelve 'intermediate'
redshift clusters ($0.11\le z\le0.32$). We also require the clusters
to have at least $\sim20,000$ ACIS-S or ACIS-I counts in order to
study their properties in at least 3 circular annuli.  The sample is
presented in Table~\ref{sample}, where the name of the cluster and the
Chandra observing logs are listed.\\
\begin{deluxetable*}{lccccccc}
\tabletypesize{\footnotesize}
\tablecaption{Observation Log for the Chandra Cluster Archive Sample. 
Column (2) shows the redshift $z$ of the clusters. The columns
(3), (4) and (5) show the instrument used, the observation date and
the observation ID, respectively. The columns (6) and (7) show
the observing time before ($t_{exp}$) and after ($t_{clean}$)
the removal of high background intervals. Column (9) is the Galactic
column density $N_H$ in the line of sight of the observation.
\label{sample}}
\tablewidth{0pt}
\tablehead{
\colhead{Name} &
\colhead{$z$} &
\colhead{Instrument} &
\colhead{Obs. Date} &
\colhead{Obs. ID} &
\colhead{\begin{tabular}{c}
$t_{exp}$\\
(ksec)
\end{tabular}} &
\colhead{\begin{tabular}{c}
$t_{clean}$\\
(ksec)
\end{tabular}} &
\colhead{\begin{tabular}{c}
$N_H$\\
(10$^{20}$ cm$^{-2}$)
\end{tabular}}
}
\startdata
A2034    & 0.113 & ACIS-I & 2001 May 05 & 2204 & 53.9 & 53.9 & 1.6 \\
A1413    & 0.143 & ACIS-I & 2001 May 16 & 1661 &  9.7 &  9.7 & 2.2 \\
         &       & ACIS-I & 2004 Mar 06 & 5003 & 76.1 & 75.0 &     \\
         &       & ACIS-I & 2005 Feb 03 & 5002 & 37.2 & 36.5 &     \\
A907     & 0.153 & ACIS-I & 2000 Jun 29 &  535 & 11.0 & 10.9 & 5.4 \\
         &       & ACIS-I & 2002 Jun 14 & 3185 & 48.7 & 47.9 &     \\
         &       & ACIS-I & 2002 Oct 30 & 3205 & 47.7 & 40.5 &     \\
A2104    & 0.155 & ACIS-S & 2000 May 25 &  895 & 49.8 & 48.9 & 8.7 \\
A1914    & 0.171 & ACIS-I & 2003 Sep 03 & 3593 & 18.9 & 18.8 & 0.9 \\
A2218    & 0.176 & ACIS-I & 2001 Aug 30 & 1666 & 49.2 & 20.2 & 3.2 \\
A963     & 0.206 & ACIS-S & 2000 Oct 11 &  903 & 36.8 & 35.8 & 1.4 \\
A2261    & 0.224 & ACIS-I & 2004 Jan 14 & 5007 & 24.6 & 24.3 & 3.3 \\
A2390    & 0.228 & ACIS-S & 2000 Oct 08 &  500 &  9.8 &  9.8 & 6.8 \\
         &       & ACIS-S & 2003 Sep 11 & 4193 & 96.3 & 91.0 &     \\
A1835    & 0.253 & ACIS-S & 2000 Apr 29 &  496 & 10.8 & 10.3 & 2.3 \\
ZwCl3146 & 0.291 & ACIS-I & 2000 May 10 &  909 & 46.6 & 45.6 & 3.0 \\
A1995    & 0.319 & ACIS-S & 2000 May 08 &  906 & 57.5 & 53.8 & 1.4 \\
\enddata

\end{deluxetable*}
The Chandra data analysis has been performed using the latest version
of CIAO (v3.3.0.1).  All of our datasets are processed by a version of
the Standard Data Processing (SDP) pipeline prior to version DS 7.4.0,
which uses the tool $acis\_detect\_afterglow$ to flag possible cosmic
ray events in the level 1 event file; it has been determined that a
significant fraction of the X-ray events from a source in imaging mode
might be removed using this tool.  Therefore we reset the correction
performed by $acis\_detect\_afterglow$ on the Level=1 event file, so
that the hot pixels and the afterglow events may be properly removed
by the improved CIAO tool $acis\_run\_hotpix$, (introduced after SDP
version DS 7.4.0). A new Level=1 event file is then created (through
the CIAO tool $acis\_process\_events$) to apply the latest calibration
files to the data (e.g.  apply the newest ACIS gain maps, apply the
time-dependent ACIS gain correction, apply the ACIS Charge Transfer
Inefficiency correction, etc.). Moreover in the case of observations
telemetered in VFAINT mode it is possible to reduce the background
using the additional screening of the events with significantly
positive pixels at the border of the $5\times5$ event island.  Two
further filtering steps are then required to obtain the Level=2 event
files, i.e. filter for bad grades (using ASCA grades) and for a
"clean" status column and apply the Good Time Intervals (GTIs)
supplied by the pipeline.  The final step is to examine background
light curves during each observation to detect and remove the periods
of high background, due to flaring episodes. We perform the flare
detection and removal following the recommendations suggested in
Markevitch et al. (2003); both the total and the clean exposure times
are listed in Table~\ref{sample}. Most of the observations are
slightly affected by background flares thus we were able to use
practically all the exposure time. The only exceptions are ObsID 3205,
4193, 906 and especially ObsID 1666, where $\sim29$ ks of the exposure
were lost due to high background.\\

\subsection{Background Subtraction}
An accurate subtraction of the background is crucial to perform a
correct study of the spectral properties of the clusters in our
sample, especially in their outskirts. Since we are dealing with
extended objects, occupying most of the ACIS field of view we need to
use a compilation of the blank-field observations, processed
identically to the cluster observation (i.e. as described above) and
reprojected onto the sky using the aspect information from the cluster
pointing. It is worth noticing that the synthetic backgrounds
correspond to longer exposure times ($\sim0.5$ Msec) than any of our
observations, giving us a very good sampling in the estimate of the
background to subtract.  Moreover, in order to 'tailor' the background
to our data we follow the recommendations given in the CIAO
web-pages\footnote{{\tiny http://cxc.harvard.edu/cal/Acis/Cal\_prods/bkgrnd/acisbg/COOKBOOK}}.
In particular we renormalize the blank-fields to the background in
each observation, considering a region of the ACIS field of view
practically free from cluster emission (mainly ACIS-S1 for ACIS-S
observations, and ACIS-S2 for ACIS-I observations) and a spectral band
(9.5-12 keV) where the Chandra effective area is nearly zero,
therefore all the observed flux is due to the particle background.

In addition to the particle-induced background we check also if the
diffuse soft X-ray background could be an important factor
in our observations 
and if appropriate adjustments are needed. For each observation, 
we follow the procedure of Vikhlinin et al. (2005), extracting a 
spectra in the source-free regions of the detector, subtracting the
renormalized blank-field background and fitting the residuals in
XSPEC v11.3.2p (in the 0.4-1 keV band) with an unabsorbed {\tt mekal}
model, whose normalization was allowed to be negative.
The best-fit model obtained is therefore included as an additional 
component in the spectral fits (with its normalization scaled by the area).
However, in every observation the adjustments required are minimal 
and do not affect significantly the determination of $kT$ and $Z$, even at large radii. 
This is also due to the properties of the clusters in our sample, 
whose high $kT$ values ($>4$ keV) even in the resolved outer regions are not
affected significantly from the method applied for the subtraction of the diffuse
soft background.

\subsection{Cash statistics vs. $\chi^2$ statistics}
The $\chi^2$ statistics require grouping of the spectra, having at least
20 counts per bin, in order to be able to approximate the Poissonian
distribution of counts with a Gaussian. On the contrary, Cash
statistics do not require any grouping and represent a more reliable
(and less biased) approach to fit the data.  Indeed, it is well known
from the literature (see e.g. Nousek \& Shue 1989; Balestra et
al. 2007)
that the $\chi^2$ statistics systematically ``sees'' the observed
spectra softer than the real ones.  This usually leads to an
overestimate of the slope of the observed spectra in the case of a
simple power-law fit, while in the case that a thermal model is fitted
to the data, the temperature measured is usually underestimated.
As a test to see whether this systematics is present also in our data,
we have decided to apply both these fit statistics.
We find that for all the clusters of the sample, a systematically
lower temperature is measured with the $\chi^2$ (on average $7\%-17\%$
lower, depending on the cluster). On the other hand no obvious
systematic trend is observed in the determination of $Z$, being the
variation in the best--fitting value of the metallicity in each
cluster ranging between $\Delta Z\sim0.01$ and $\sim0.07$
with no preferential direction.
To avoid the dependence on the grouping method, and the bias in the
best-fit temperature, we decide to use the modified Cash statistics, 
as implemented in XSPEC v11.3.2p,
to determine the best-fit parameters and their uncertainties.

\subsection{Spectral Analysis}\label{specanal}
In order to study the radial properties of the cluster emission, we
subdivide each cluster in annuli (circular or elliptical, depending on
the morphology of the cluster) centered on the X-ray emission peak. In
the more disturbed clusters, where an emission peak is not clearly
identifiable, we assume the center of the cluster to correspond with
the X-ray centroid at $0.5r_{500}$.  We require each region to have at
least $\sim7,000$ net counts, so that it would be possible to estimate
the temperature and the metallicity of the annulus with sufficient
accuracy.
For each cluster the outermost annulus corresponds to an area where
the intensity of the source in the 0.8-8 keV band is roughly equal to
that of the background.  We extract a spectra from each annulus after
excluding the 3$\sigma$ point sources detected by the CIAO tool
$wavdetect$. The source list produced is also inspected 'by-eye' in
order to remove possible additional sources not detected by
$wavdetect$ (especially in the regions where the diffuse emission from
the cluster is brighter).  The CIAO script used to perform the
spectral extraction is
$specextract$, which generates source and background spectra and build
the appropriate RMFs and ARFs. The background
is taken from the re-normalized blank field observations using the same region of the source.\\
The spectra are analyzed with XSPEC v11.3.2p (Arnaud et al. 1996) and
fitted by a single-temperature {\tt mekal} model (Kaastra 1992;
Liedahl et al. 1995) in which the ratio between the elements is fixed
to the solar value as in AG89. However, as explained in \S1, these
values for the solar metallicities have more recently been superseeded
by the new values by Grevesse \& Sauval (1998) and A05. For clarity
and completeness, we also performed the fits using solar abundances by
A05.  The free parameters in the model are the temperature $kT$, the
metallicity $Z$ of the gas and the normalization.  The spectral band
considered in the fit is the 0.6-8 keV. We choose not to consider the
data below 0.6 keV because of uncertainties in the ACIS calibration
below that energy.
The $N_H$ derived from the X-rays is
found to be
consistent (within 1$\sigma$) with the Galactic value in the line of
sight of each observation, as derived from radio data (Stark et
al. 1992), except in the cases of A2104 and A2390 (see
Table~\ref{sample} and following \S~\ref{text2104}). In these
clusters the $N_H$ value measured from X-ray data is significantly
different (at more than 2$\sigma$ confidence level) from the radio
value, therefore we adopt the X-ray value. The $N_H$ value is fixed to
the Galactic value obtained from the radio data (and listed
in Table~\ref{sample}) in the rest of the sample.
We have measured $N_H$ from the X-ray data in each annulus, finding no
evidence of radial variation. Therefore, $N_H$ is fixed to the same value
in all the radial annuli. 

\begin{figure}
\plotone{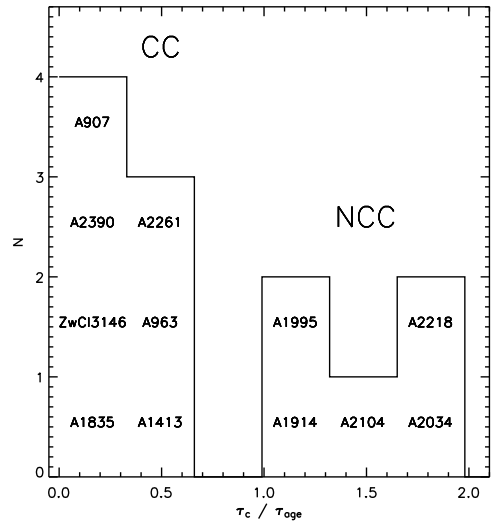}
\caption{Histograms of the distribution in our sample of the values of 
$\tau_c / \tau_{age}$, computed as described in \S~\ref{selftemp} and quoted 
in Table~\ref{xrayprop}.}
\label{histotc}
\end{figure}

\begin{figure*}
\plotone{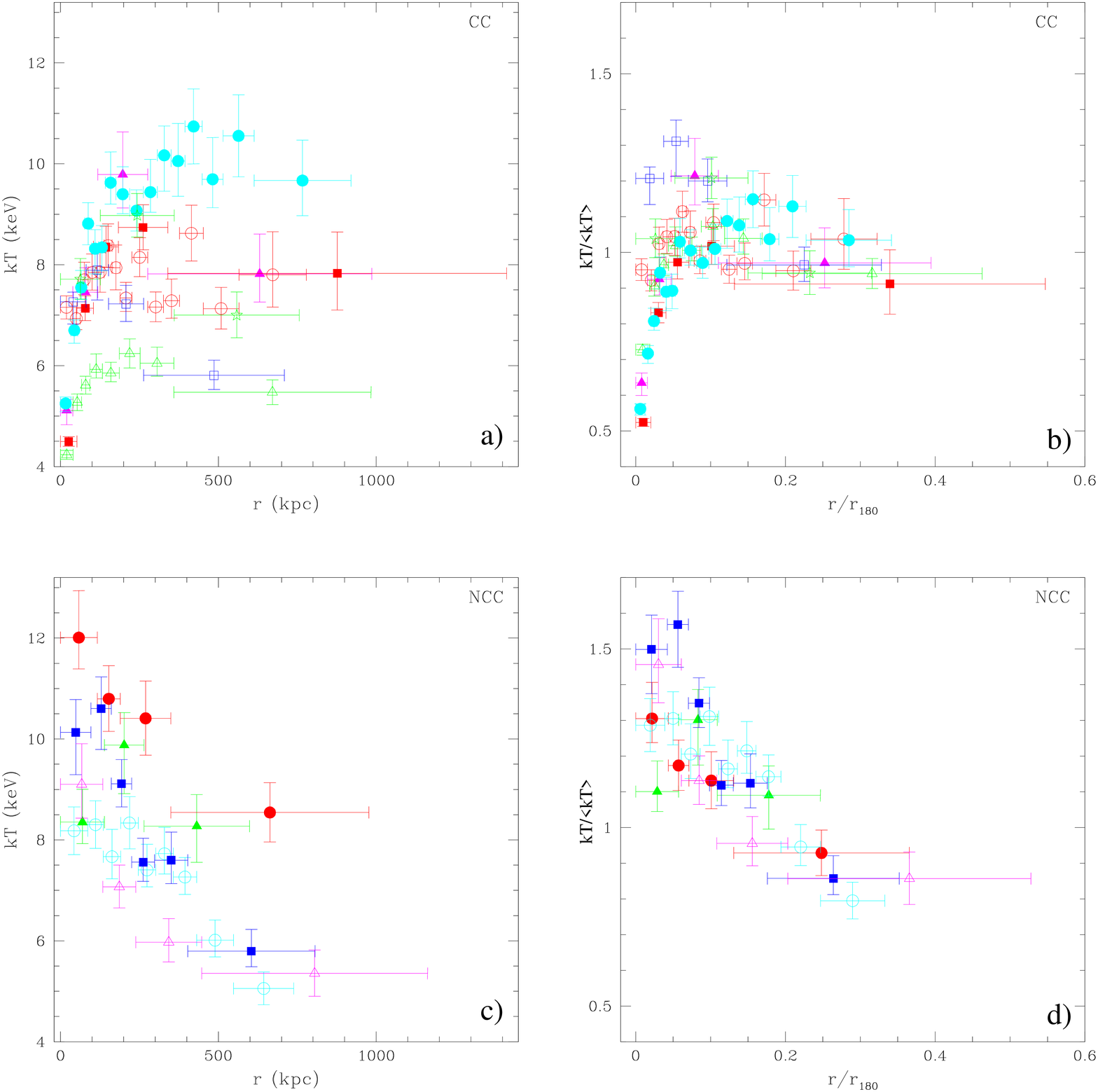}
\caption{{\it a):} Temperature profiles for the CC clusters in the sample: A907 (green empty triangles), 
A963 (blue empty squares), A1413 (red empty circles), 
A1835 (magenta filled triangles), A2261 (green stars), 
A2390 (cyan filled circles) and ZwCl3146 (red filled squares).
{\it b):} Normalized temperature profiles for the CC clusters, plotted
against the radii in units of $r_{180}$. The symbols have the same meaning as in panel a).
{\it c):} Temperature profiles for the NCC clusters in the sample: A1914 (red filled circles), 
A1995 (green filled triangles), A2034 (cyan empty circles), 
A2104 (blue filled squares) and A2218 (magenta empty triangles).
{\it d):} Normalized temperature profiles for the NCC clusters, plotted
against the radii in units of $r_{180}$. The symbols have the same meaning as in panel c).
}
\label{temperatures}
\end{figure*}

\begin{figure*}
\plotone{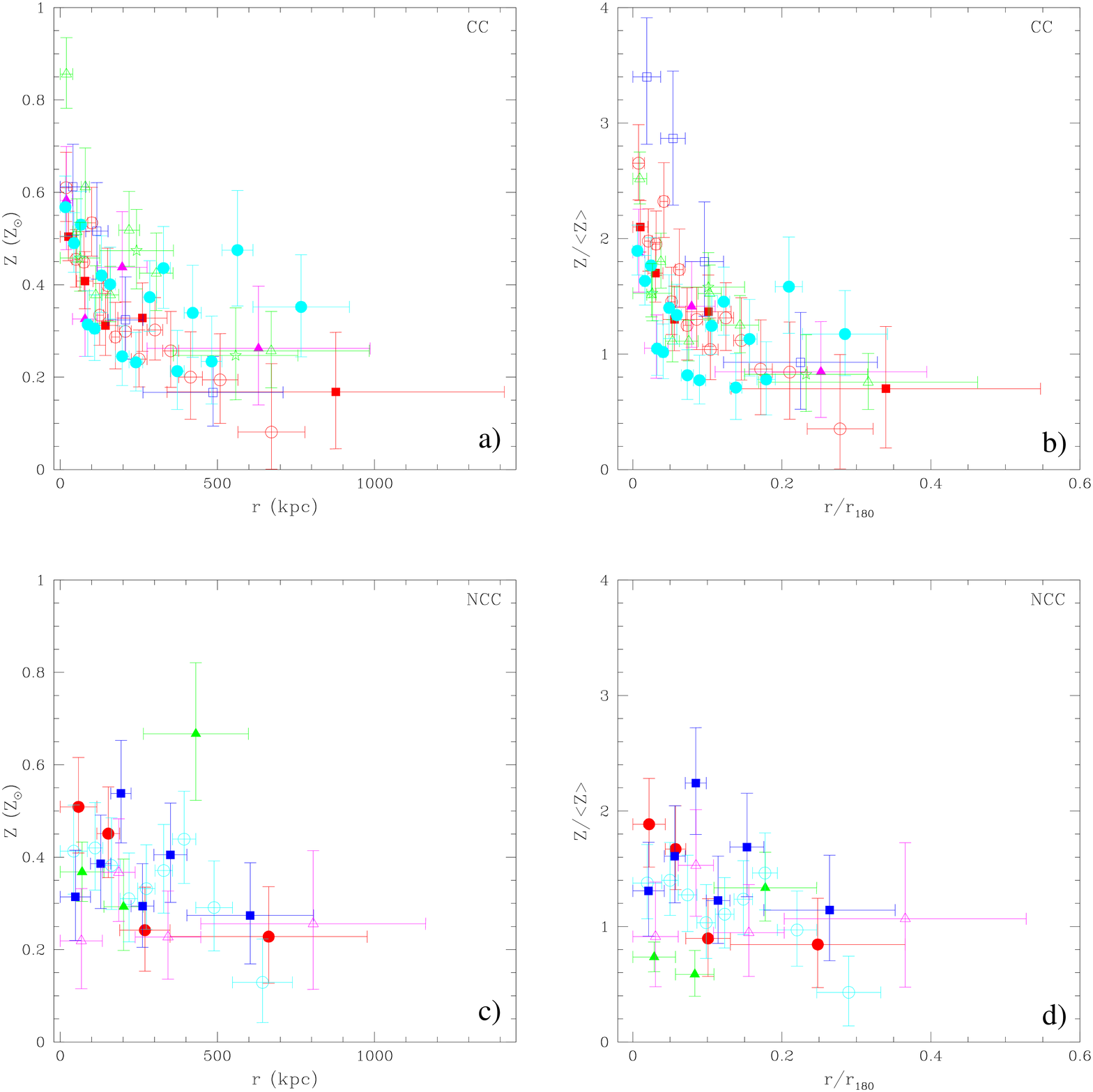}
\caption{{\it a):} Abundance profiles for the CC clusters in the sample: the different symbols correspond
to the clusters listed in Figure~\ref{temperatures}a.
{\it b):} Normalized abundance profiles for the CC clusters, plotted
against the radii in units of $r_{180}$. The symbols have the same meaning as in panel a).
{\it c):} Abundance profiles for the NCC clusters in the sample: the different symbols correspond
to the clusters listed in Figure~\ref{temperatures}c.
{\it d):} Normalized abundance profiles for the NCC clusters, plotted
against the radii in units of $r_{180}$. The symbols have the same meaning as in panel c)}
\label{abundances}
\end{figure*}

We have divided then our sample in Cooling-Core and
Non-Cooling-Core clusters according to their central cooling time.
The gas temperature and density profiles are recovered from the single-phase 
spectral fit done in annular rings by correcting the emissivity in each shell 
by the contrubution of the outer shells moving inwards. A detailed description 
of the procedure is presented in Ettori et al. (2002). In brief, the 
normalization of the thermal component, being proportional to the Emission 
Integral, provides the gas density, whereas the deprojected temperature is 
provided by weighting for the corrected emissivity the spectral measurement.  
The deprojected values in the innermost bin are then used to estimate the 
central cooling times $\tau_c =
5/2 (\mu_e/\mu) T_e (n_e / \epsilon)$, where $\mu=0.613$ and
$\mu_e=1.174$ are appropriate for a plasma with a metallicity of 0.3
times the solar values in AG89, $T_e$, $n_e$ and $\epsilon$
are the gas temperature, electron density and emissivity in the
innermost bin, respectively.\\
\begin{deluxetable*}{lccccccc}
\tabletypesize{\small}
\tablecaption{Global cluster properties of the Chandra sample. 
Column (2) shows the the total net counts from the inner to the outer annulus 
considered in the spectral analysis. The columns (3), (4) and (5) show the 
global temperature $\langle kT \rangle$, the global metallicity 
$\langle Z \rangle$ and the virial radius $r_{180}$, computed within
$0.07-0.4r_{180}$, respectively. Column (6) shows the aperture used to measure 
$\langle kT \rangle$ and $\langle Z \rangle$.
The columns (7) and (8) show the central cooling times and the ratio with 
respect to the age of the Universe at the cluster redshift.
\label{xrayprop}}
\tablewidth{0pt}
\tablehead{
\colhead{Name} &
\colhead{\begin{tabular}{c}
Net cts\\
(0.6-8 keV)
\end{tabular}} &
\colhead{\begin{tabular}{c}
$\langle kT \rangle$\\
(keV)
\end{tabular}} &
\colhead{\begin{tabular}{c}
$\langle Z \rangle$\\
($Z_\odot$)
\end{tabular}} &
\colhead{\begin{tabular}{c}
$r_{180}$\\
(kpc)
\end{tabular}} &
\colhead{\begin{tabular}{c}
Aperture\\
($^{\prime\prime}$)
\end{tabular}} &
\colhead{$\tau_c$ (Gyr)} &
\colhead{$\tau_c/\tau_{age}$}
}
\startdata
A2034    & 78,900  & $6.36\pm0.15$  & $0.30\pm0.04$ & 2,222 & 76-433 
& $21.2\pm3.2$ & $1.77\pm0.27$\\
A1413    & 181,500 & $7.52_{-0.12}^{+0.20}$  & $0.23\pm0.03$ & 2,416 & 67-385 
& $4.2\pm0.3$  & $0.36\pm0.03$\\
A907     & 87,500  & $5.82\pm0.12$  & $0.34\pm0.04$ & 2,125 & 56-320 
& $2.0\pm0.1$  & $0.17\pm0.01$\\
A2104    & 63,100  & $6.76\pm0.19$  & $0.24\pm0.05$ & 2,290 & 60-341 
& $18.1\pm2.1$ & $1.58\pm0.18$\\
A1914    & 39,100  & $9.20_{-0.37}^{+0.39}$ & $0.27\pm0.07$ & 2,672 & 64-367 
& $12.2\pm1.0$ & $1.07\pm0.09$\\
A2218    & 18,300  & $6.25\pm0.31$  & $0.24\pm0.07$ & 2,202 & 52-295 
& $21.3\pm1.8$ & $1.89\pm0.16$ \\
A963     & 41,800  & $6.02_{-0.19}^{+0.28}$  & $0.18\pm0.06$ & 2,161 & 45-256 
& $6.5\pm0.4$ & $0.59\pm0.04$ \\
A2261    & 21,500  & $7.43_{-0.27}^{+0.49}$  & $0.30_{-0.06}^{+0.07}$ & 2,400 & 47-267
& $7.0\pm0.4$  & $0.65\pm0.03$ \\
A2390    & 202,600 & $9.35\pm0.15$ & $0.30\pm0.03$ & 2,693 & 52-295
& $1.3\pm0.2$  & $0.12\pm0.02$\\
A1835    & 23,100  & $8.06\pm0.53$  & $0.31\pm0.09$ & 2,500 & 44-254
& $0.9\pm0.1$  & $0.08\pm0.01$\\
ZwCl3146 & 40,500  & $8.59\pm0.39$  & $0.24\pm0.06$ & 2,582 & 41-237
& $1.0\pm0.1$  & $0.10\pm0.01$\\
A1995    & 30,200  & $7.59_{-0.44}^{+0.57}$  & $0.50_{-0.11}^{+0.12}$ & 2,427 & 37-209
& $12.7\pm1.3$ & $1.28\pm0.14$ \\
\enddata
\end{deluxetable*}
The central cooling times are reported in Table~\ref{xrayprop}, as well as
the age $\tau_{age}$ of each cluster, and the
ratio between the two quantities. The age of the Universe at the $z$ of
observation is used as an upper limit to the age of the cluster.
Bauer et al. (2005) computed the
cooling times for 6 of the clusters in our sample (A1835, A1914,
A2218, A2261, A2390 and ZwCl3146) finding a $\tau_c$ in the center of
the cluster, or at 50~kpc, consistent with the values computed for the
central bin in our spectral analysis (which might extend farther out
than 50~kpc from the center in some cases).  Following their
criterium, a clear separation between the CC and the NCC in our sample
can be located at $\tau_c\sim10$ Gyr (Figure~\ref{histotc};
corresponding to $\tau_c/\tau_{age}\sim1$).  We have 4 clusters
presenting signs of strong cooling ($\tau_c<2$ Gyr) and 3 clusters
exhibiting signs of mild cooling ($\tau_c<10$ Gyr). The remaining 5
clusters can be classified as NCC,
presenting longer cooling times in the center.
The projected temperature and metal abundance profiles for 
both CC and NCC objects are shown in Figures~\ref{temperatures} and
~\ref{abundances}.

\subsection{Notes on individual clusters} \label{individual}

{\bf A2034}:
A2034 (z=0.113) has been observed with Chandra in one ACIS-I pointing 
(ObsID: 2204).
The temperature profile we derived is quite flat in the central
regions of the cluster, where the temperature is $kT\sim8$ keV. It
shows however a negative gradient after 400 kpc from the center.  A
similar trend is observed in the metallicity, where the average value
of $Z=0.4$ within 400 kpc from the center decreases to $Z<0.2$ at
larger radii.

{\bf A1413}:
A1413 (z=0.143) has been observed four times with ACIS-I. We discard one 
observation (ObsID: 537) which is affected almost entirely by a
persisting flare.
In one of the observations used in our analysis (ObsID: 5003) 
the source is placed in a position of the ACIS-I array very close to
the S2 chip, therefore S2 is still contaminated by source emission and
we can not use it to re-normalize the blank field to the background in
the observation. We use instead part of the I1 chip (which is
front-illuminated as S2) to re-normalize, since it is more distant
from the cluster center than S2 and therefore
less contaminated by cluster emission.
The resulting temperature profile shows a slight decrease in
temperature towards the center ($\Delta kT=1.2_{-0.6}^{+0.7}$ keV within the
inner 150 kpc).  A1413 has been also observed by XMM-Newton (Pratt \&
Arnaud 2002), representing one of the clusters with the most accurate
temperature profile observed by this satellite. This cluster is also
part of the sample of Chandra clusters analyzied by Vikhlinin et
al. (2005).  The XMM-Newton observation does not find any evidence of
a cool core, in contrast with the temperature profiles obtained with
Chandra both in our analysis and even more evidently in Vikhlinin et
al. (2005). This might be due to the poorer angular resolution of
XMM-Newton with respect to Chandra.\\
The metallicity profile is decreasing towards larger radii, and
consistent within 1$\sigma$ with the measures of Vikhlinin et
al. (2005).

{\bf A907}:
A907(z=0.153) has been observed with Chandra in three separate ACIS-I
pointings (ObsID: 535, 3185 and 3205), all of them used
in our analysis.
The temperature profile shows an evidence of a cool core in the center
of the cluster ($\Delta kT=1.4_{-0.3}^{+0.2}$ keV in the central 100 kpc). The
metallicity profile presents a decreasing trend toward larger radii.
A907 is also part of the cluster sample analyzed by Vikhlinin et
al. (2005). Their results, both for the temperature and the
metallicity, are fully consistent with ours within the 1$\sigma$
statistical uncertainties.
 
{\bf A2104}:\label{text2104}
A2104 (z=0.155) has been observed with Chandra in one ACIS-S pointing
(ObsID: 895).
As described in \S~\ref{specanal}, the value of the $N_H$ measured
from the X-ray data alone is significantly different from the radio
value, thus we have decided to fix the $N_H$ to the best fit
value obtained from the fit ($1.55\times10^{21}$ cm$^{-2}$).
The cluster does not show any evidence of a cool core in its center,
having a temperature profile decreasing towards the outskirts.  The
metallicity profile is consistent with being flat, with a value
$Z\sim0.3-0.4Z_\odot$, within the 1$\sigma$ uncertainties.

{\bf A1914}:
Two ACIS-I pointings of A1914 (z=0.171) are available in Chandra
archive. However the oldest (and shortest) observation (ObsID: 542)
has been performed in 1999. For observations performed in that year an
accurate modeling of the ACIS background is not currently
available. Thus to avoid problems in background subtraction we have
decided to discard it and keep only the longest observation
(ObsID: 3593).
A negative gradient in $kT$ is quite clear: the temperature drops from
$kT=12.0_{-0.6}^{+0.9}$ keV in the center down to 
$kT=8.5\pm0.6$ keV in the outer radial
bin. A similar trend is observed also in the abundance profile, where
$Z=0.5\pm0.1Z_\odot$ in the center, then decreasing to 
$0.2\pm0.1 Z_\odot$ in the
two outer radial bins.
This is one of the few examples of metallicity peak without a
corresponding cool core (or temperature drop) towards the center.

{\bf A2218}:
A2218 (z=0.176) has been observed three times with
ACIS-S. Unfortunately two of these observations (ObsID: 553 and 1454)
were performed in 1999 and for the reason described in the case of
A1914 we have decided to discard them. Moreover the remaining
observation (ObsID: 1666) has been strongly affected by a flare which
reduces the good exposure time to
only $\sim20$ ks.
With these data we are able to observe the presence of a centrally
peaked temperature profile (a hot, instead of a cool core) and a costant
metallicity profile.  A temperature profile peaked toward the center
has been also seen by Machacek et al. (2002), analyzing the two
Chandra observations performed in 1999.  This is consistent with the
picture of A2218 being involved in a line-of-sight merger, as
suggested by a considerable disturbance of the intracluster gas in the
X-rays and by the observed substructure in the optical (e.g. Pratt et
al. 2005).

{\bf A963}:
A963 (z=0.206) has been observed with Chandra in one ACIS-S pointing (ObsID: 903).
We found a decreasing trend of $Z$ with the radius, with only a very weak hint for the presence
of a lower temperature in the center. 

{\bf A2261}:
Two pointings of A2261 (z=0.224) are available in the Chandra archive. One of the observations
(ObsID: 550) has been performed in 1999 and therefore we discard it for the reason described 
above in the case of A1914.
The temperature profile does show only a hint (more than 2$\sigma$
however)
of a decrease in the center, where 
the temperature drops down from $9.0\pm0.4$ keV to $7.7\pm0.4$ keV. 
The metallicity profile shows a constant behaviour for the first two bins and a decrease (significant
at more than 1$\sigma$) in the outer radial bin.

{\bf A2390}:
A2390 (z=0.228) has been observed three times with ACIS-S. One of the observations (ObsID: 501) has
been performed
in 1999 and therefore we discard it. We concentrate our analysis on the remaining two observations
(ObsID: 500 and 4193), yielding a total of $\sim100$ ks of good observing time.
The value of the $N_H$ derived from the X-rays ($N_H=1.1\times10^{21}$
cm$^{-2}$) is significantly different than the radio value, therefore we adopted 
the X-ray value in the spectral fits.
A2390 is also part of the sample analyzed in Vikhlinin et al. (2005). Similarly to them we find a cool
core ($kT=5.8\pm0.2$ keV) in the center of the cluster with a $kT$ profile getting flatter going towards the
outskirts, being fully consistent with their measured temperatures at every radius. 
On the other hand, the metallicity profile shows just a hint of a peak in the central part of the 
cluster. It is however consistent at 1$\sigma$ with the profile of Vikhlinin et al. (2005) and not sensitive to
the choice of the $N_H$.

{\bf A1835}:
Two different ACIS-S observations of A1835 (z=0.253) are present in
the Chandra archive. We discard the older (and longer) observation
(ObsID: 495), performed in 1999 because of the reason described
in the case of A1914, keeping only the $\sim10$ ks observation performed in 2000 (ObsID: 496).
The temperature profile of A1835 shows a clear evidence of a cool core
in its center where $kT$ drops down by a factor of $\sim2$.  Moreover
the temperature shows a decline after 300 kpc, going toward larger
radii.  Piffaretti et al. (2005) analysed XMM-Newton observations of
A1835 and detected a temperature decrease at large radii, as in our
data. Majerowicz et al. (2002) also analysed XMM-Newton data and found
a decrease in the temperature profile at large radii (at $\sim400$ kpc
from the center); however their temperature profile becomes constant
after such decrease.  The decrease at large radii has not been
observed in the analysis of Chandra data by Voigt \& Fabian (2006),who
found a constant temperature outside the central 100 kpc. However it
is worth noticing that in their work they analysed the 1999
observation (instead of the 2000 observation, as in our analysis) which
might have background subtraction problems especially at large
radii. This may explain the difference between the
two profiles.
The metallicity profile shows a decreasing gradient in the first two
bins, becoming constant afterwards.  The only comparison with the
literature comes from an XMM-Newton observation analyzed by Majerowicz
et al.  (2002), where an almost constant metallicity profile at every
radius has been observed.

{\bf ZwCl3146}:
ZwCl3146 (z=0.291) has been observed with ACIS-I in one pointing (ObsID: 909).
Also this cluster clearly shows the presence of a cool core ($kT$
dropping down by a factor of almost 2). A decreasing trend in
metallicity from $Z=0.50\pm0.05$ in the center, down to $Z=0.17\pm0.12$ in the
outer bin, is also observed.

{\bf A1995}:
One ACIS-S observation of A1995, the farthest cluster in our sample
(z=0.319), is present in the Chandra archive (ObsID: 906). Although
this observation is quite long ($\sim50$ ks of good exposure
time) the number of counts available allowed us to divide this cluster 
only in three radial bins.
The temperature profile of A1995 is consistent to be flat within the
errors with a temperature around 9 keV.  The abundance profile seems
to have a positive gradient in the outer bin, however the errors are
large and this increase in $Z$ is not very significant.

\section{Self-similarity of radial profiles} \label{selftemp}

One of the main goals of this paper is to look for a (purely
phenomenological) self-similarity in the radial profiles of
temperature and metallicity, after they are scaled to the cluster
virial radius $r_{180}$. A measure of $r_{180}$ is thus crucial to
test for such self-similarity in our cluster sample. This quantity can
be approximated by the following relation:
\begin{equation}
r_{180}=1.95h^{-1}\:Mpc(\langle kT \rangle/10\:keV)^{1/2},
\end{equation}
as calibrated from the non-radiative hydrodynamical simulations of
clusters by Evrard et al. (1996). It is worth noticing that this
relation is in agreement with the scaling relations observed
(e.g. Ettori et al. 2004) in the X-rays, where the dependency on $kT$ is
consistent with Eq.(1) and only the absolute normalization may
experience some variations. To compute the global temperature
$\langle kT \rangle$ necessary to estimate $r_{180}$ we extract spectra
including emission going from 0.07$r_{180}$ to 0.4$r_{180}$ in each
cluster.  The central regions of each cluster are therefore excluded
from the spectra in order to avoid contamination from a possible cool
core. The values of $\langle kT \rangle$ and $r_{180}$ have been
evaluated iteratively until a convergence to a stable value of the
temperature is obtained ($\Delta kT\leq0.01$ keV between two different
iterations).
From the fits we are able to determine also a global metallicity $\langle Z \rangle$ in each cluster.
In Table~\ref{xrayprop} we list the best-fit values for $\langle kT \rangle$ and $\langle Z \rangle$, 
and the value of $r_{180}$ computed using
the formula above. The minimum and maximum apertures used to extract the total spectrum are listed as well.\\

\subsection{The temperature profiles}

Figure~\ref{temperatures} shows the normalized temperature profiles
for all the CC clusters (panel b) compared with the NCC clusters
(panel d). This figure has been obtained by normalizing the temperatures
in each cluster to its average temperature $\langle kT \rangle$
computed from the total cluster spectrum excluding the central
$0.07 r_{180}$.
The error-weighted mean and the best-fit results after fitting with
single power-laws $Y \propto r^{\mu}$ are presented in Table~\ref{bestfit}.

\begin{figure}
\plotone{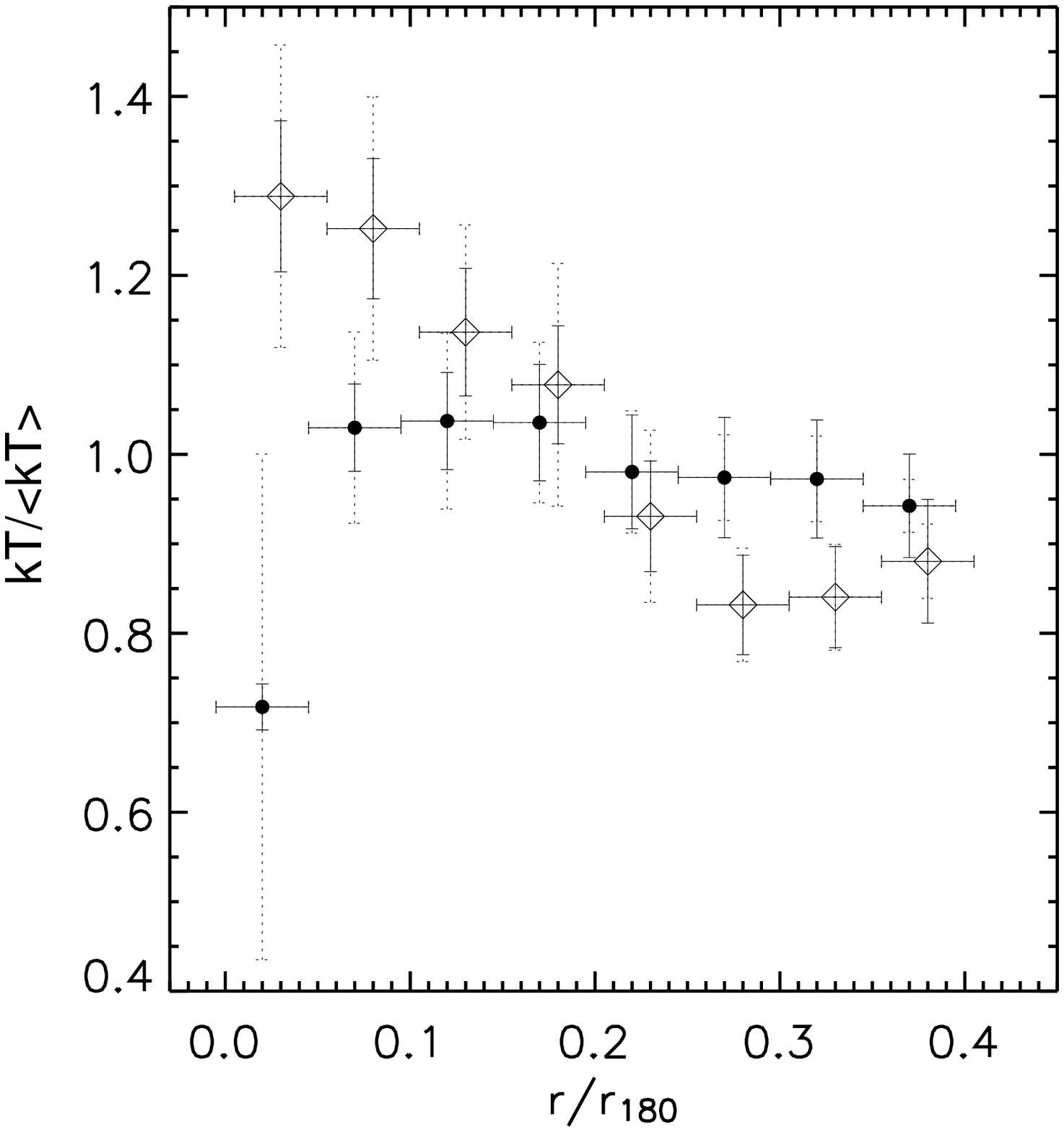}
\plotone{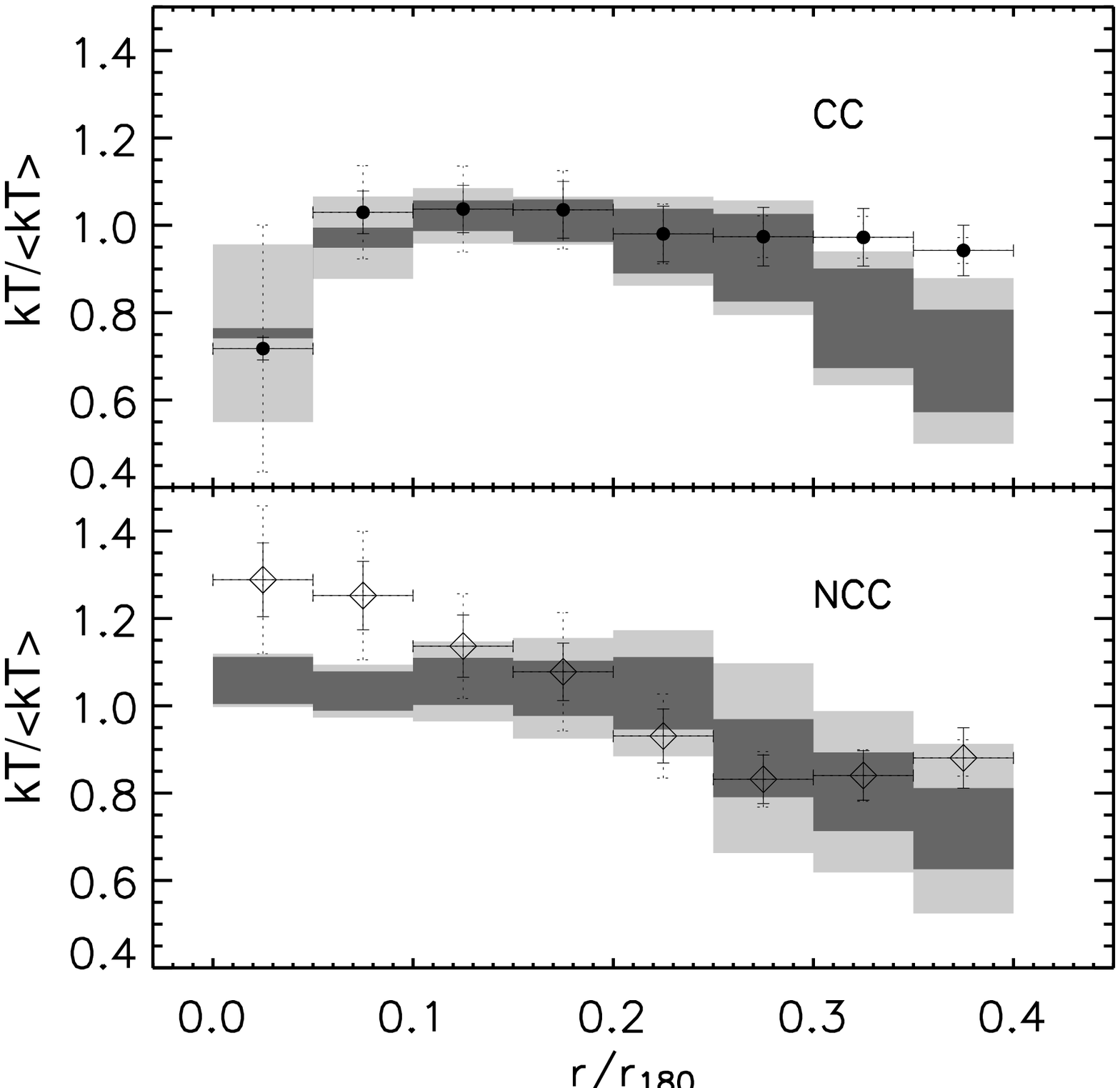}
\caption{Error-weighted mean temperature profile of the CC ({\it filled circles})
and NCC ({\it open diamonds}) clusters in our sample at intermediate redshift (top panel).
(Bottom) Comparison between our results and the local measurements in DM02 (shaded regions).
The 1$\sigma$ errors on the means are plotted as solid lines (dark gray region for the local
estimates) while the scatter (rms) in each data bin is shown as a dotted line
(light gray region for local estimates).}
\label{T_CC_vs_noCC_mean}
\end{figure}

Within $0.1 r_{180}$, the temperature profiles in CC objects increases with a slope
$\mu = 0.25$. Moving outwards, between $0.1 r_{180}$ and the outer radial limit
of our spectral analysis at $\approx 0.5 r_{180}$, these profiles behave as $r^{-0.1}$.
Non-cooling-cores systems have, on average, a profile that is almost flat
at $r<0.1 r_{180}$ and then decreases rapidly as $r^{-0.3}$.
In the outskirts, the temperature profiles of CC and NCC clusters
show a significant discrepancy between their slopes, being
NCC more deviant from the isothermal case.

Our best fit functional for the CC sample is fully consistent with
the best fit functional form found by Vikhlinin et al. (2005) in their sample
of CC clusters, at $r\la0.3r_{180}$.
The two functionals diverge significantly only above $0.3r_{180}$, where our 
profile is flatter (and therefore the value of $kT/\langle kT \rangle$ 
is higher) than Vikhlinin et al. (2005) profile.
However, only a few of our data points are located beyond that radius, 
preventing us from any statistically significant comparison between the two
samples at $r\ga0.3r_{180}$.

\begin{deluxetable*}{lccc}
\tabletypesize{\small}
\tablecaption{Error-weighted means, with errors on the mean and r.m.s. quoted within round brackets, 
and best-fit parameters of the single power-laws $Y = Y_{0.1} (x/0.1)^{\mu}$, with
$x = r/r_{180}$. $\langle kT \rangle$ and $\langle Z \rangle$ are measured in the radial range
$0.07-0.4 r_{180}$. \label{bestfit} }
\tablewidth{0pt}
\tablehead{
\colhead{} &
\colhead{} &
\colhead{$kT/\langle kT \rangle$} &
\colhead{}
}
\startdata
    &  all $r$ &  $r < 0.1 r_{180}$ & $r > 0.1 r_{180}$ \\
All & $0.84\pm0.04 (0.28)$ & $0.79\pm0.03 (0.35)$ & $1.02\pm0.06 (0.10)$ \\
 & $Y_{0.1}=1.00\pm0.01$ & $Y_{0.1}=1.20\pm0.01$ & $Y_{0.1}=1.10\pm0.02$ \\
 & $\mu=0.15\pm0.01$ & $\mu=0.26\pm0.01$ & $\mu=-0.16\pm0.03$ \\
 & $\chi^2$/dof=$1223.4/78$ & $\chi^2$/dof=$720.0/43$ & $\chi^2$/dof=$59.1/33$ \\
 &&& \\
CC  & $0.80\pm0.04 (0.24)$ & $0.76\pm0.03 (0.26)$ & $1.02\pm0.06 (0.08)$ \\
 & $Y_{0.1}=1.00\pm0.01$ & $Y_{0.1}=1.16\pm0.02$ & $Y_{0.1}=1.07\pm0.02$ \\
 & $\mu=0.17\pm0.01$ & $\mu=0.25\pm0.01$ & $\mu=-0.10\pm0.03$ \\
 & $\chi^2$/dof=$744.3/52$ & $\chi^2$/dof=$496.3/30$ & $\chi^2$/dof=$26.7/20$ \\
 &&& \\
NCC & $1.13\pm0.07 (0.20)$ & $1.27\pm0.08 (0.14)$ & $1.02\pm0.07 (0.13)$ \\
 & $Y_{0.1}=1.12\pm0.01$ & $Y_{0.1}=1.25\pm0.04$ & $Y_{0.1}=1.23\pm0.04$ \\
 & $\mu=-0.15\pm0.01$ & $\mu=-0.02\pm0.03$ & $\mu=-0.32\pm0.05$ \\
 & $\chi^2$/dof=$76.6/24$ & $\chi^2$/dof=$28.8/11$ & $\chi^2$/dof=$17.1/11$ \\
 &&& \\
 & & $Z/\langle Z \rangle$ & \\
\hline
    &  all $r$ &  $r < 0.1 r_{180}$ & $r > 0.1 r_{180}$ \\
All & $1.31\pm0.29 (0.54)$ & $1.41\pm0.26 (0.60)$ & $1.11\pm0.33 (0.32)$ \\
 & $Y_{0.1}=1.13\pm0.04$ & $Y_{0.1}=1.03\pm0.06$ & $Y_{0.1}=1.35\pm0.10$ \\
 & $\mu=-0.23\pm0.02$ & $\mu=-0.28\pm0.03$ & $\mu=-0.44\pm0.15$ \\
 & $\chi^2$/dof=$149.2/78$ & $\chi^2$/dof=$120.7/43$ & $\chi^2$/dof=$20.0/33$ \\
 &&& \\
CC  & $1.40\pm0.28 (0.58)$ & $1.53\pm0.26 (0.62)$ & $1.12\pm0.32 (0.33)$ \\
 & $Y_{0.1}=1.17\pm0.05$ & $Y_{0.1}=1.11\pm0.06$ & $Y_{0.1}=1.37\pm0.11$ \\
 & $\mu=-0.24\pm0.03$ & $\mu=-0.27\pm0.03$ & $\mu=-0.52\pm0.18$ \\
 & $\chi^2$/dof=$80.2/52$ & $\chi^2$/dof=$64.9/30$ & $\chi^2$/dof=$11.0/20$ \\
 &&& \\
NCC & $1.06\pm0.31 (0.44)$ & $1.05\pm0.28 (0.56)$ & $1.08\pm0.36 (0.31)$ \\
 & $Y_{0.1}=1.06\pm0.07$ & $Y_{0.1}=1.08\pm0.15$ & $Y_{0.1}=1.27\pm0.20$ \\
 & $\mu=0.00\pm0.07$ & $\mu=0.03\pm0.14$ & $\mu=-0.29\pm0.27$ \\
 & $\chi^2$/dof=$41.7/24$ & $\chi^2$/dof=$31.7/11$ & $\chi^2$/dof=$8.5/11$ \\
\enddata
\end{deluxetable*}
Adopting the A05 abundances has not changed the best-fit values of $kT$
at every radius (always fully consistent within the 1$\sigma$ errors)
in the individual clusters, therefore the best fit functionals
representing both the CC and the NCC sample
have not varied.

A clearer picture can be seen also if we compute an error-weighted
average of the $kT$ profiles in several bins of width 0.05 in $r/r_{180}$.
The contribution to the single bin is provided from the measurements (and relative
error) that fall into that bin, weighted in proportion to the percentage
of the spatial coverage of the bin.
The error-weighted mean $kT/\langle kT \rangle$ profile is plotted in
Figure~\ref{T_CC_vs_noCC_mean} and compared with the local estimates
from DM02.
CC and NCC objects show well defined opposite gradient in the inner radial
and more similar behaviour moving outwards.
A good agreement is also observed with the local profiles, apart from two
significant deviations: (i) our CC mean profile appears flatter
at $r>0.2 r_{180}$, with an error-weighted value of $0.97 \pm 0.06$
(r.m.s. 0.07), to be compared with the local value of $0.85 \pm 0.11$
(r.m.s. 0.16), (ii) our NCC profile is steeper within $0.1 r_{180}$,
with a mean value of $1.27 \pm 0.08$ (r.m.s. 0.14) with respect to
the local value of $1.04 \pm 0.05$ (r.m.s. 0.07).

\subsection{The metal abundance profiles}

The metallicity profiles are plotted against the radius
normalized to $r_{180}$ for the CC and NCC sample
in Figure~\ref{abundances}b and \ref{abundances}d, respectively.
A different behaviour
in the very central regions between the two samples is quite clear.
To characterize this behaviour, we have fitted
the normalized profiles $Z/\langle Z \rangle$ with respect
to $r/r_{180}$ with single power-laws $Y \propto r^{\mu}$
over different radial ranges.
While the NCC clusters presents a flat profile within $\sim0.1 r_{180}$,
a sharper negative gradient is observed in the CC
cluster sample ($\mu=-0.24$; see Table~\ref{bestfit}).
Also at $r > 0.1 r_{180}$, where a model with a single power-law
well reproduce the data (reduced $\chi^2$ less than 1),
the CC clusters show hints of a
steeper profile ($\mu = -0.52 \pm 0.18$ in the CC sample,
$\mu = -0.29 \pm 0.27$ in the NCC sample).

\begin{figure}
\plotone{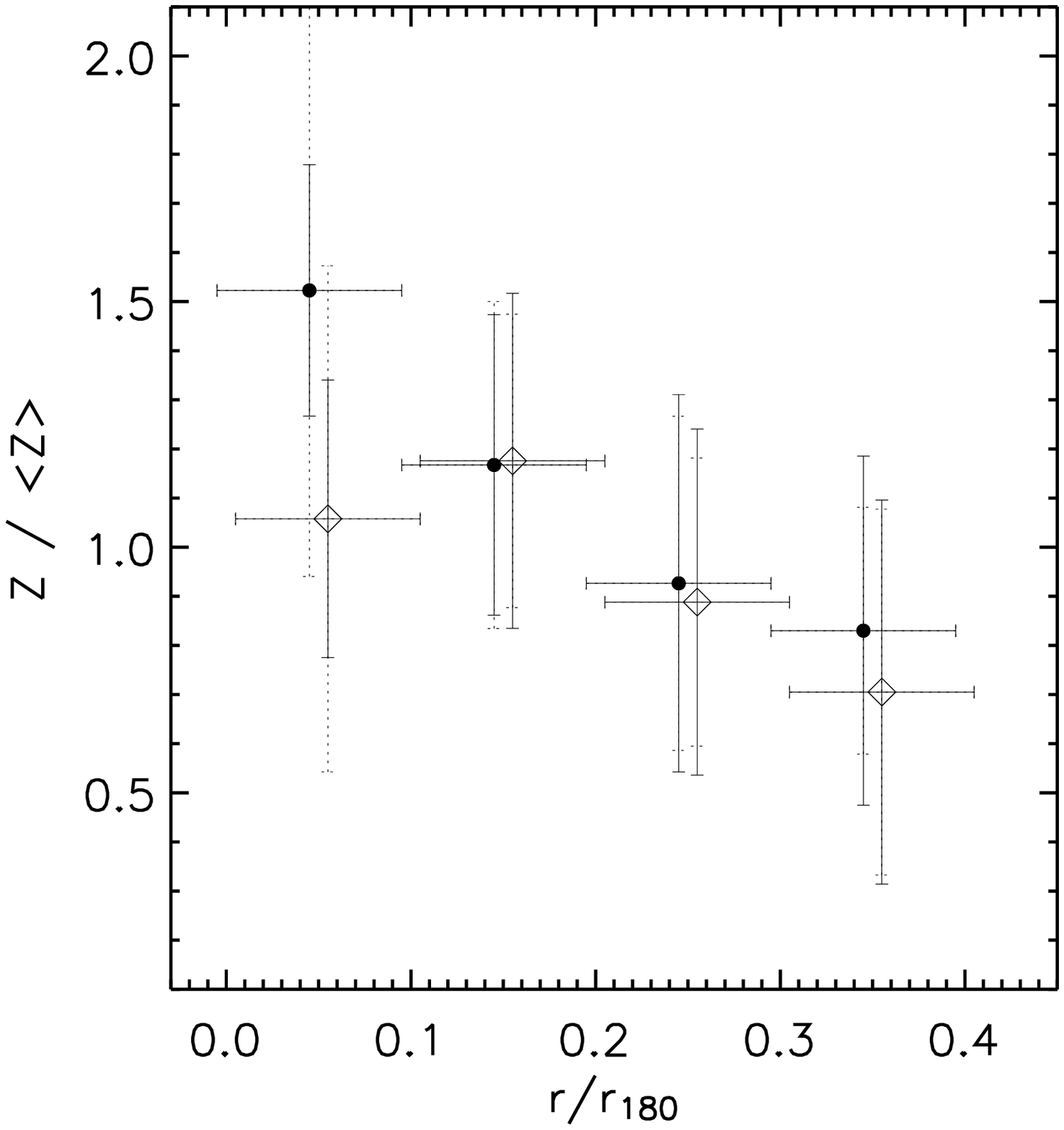}
\plotone{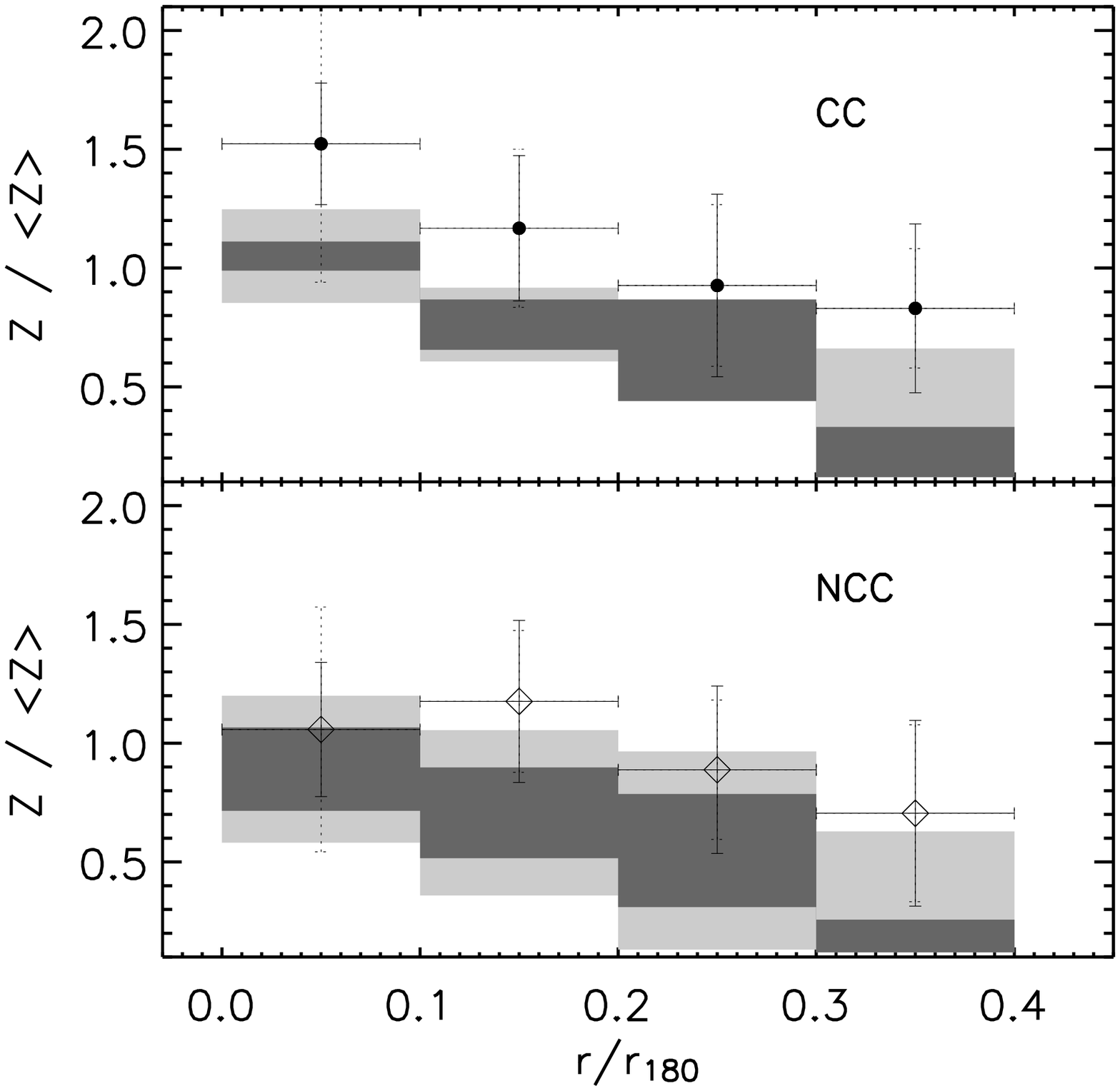}
\caption{Error-weighted mean metal abundance profile of the CC ({\it filled circles})
and NCC ({\it open diamonds}) clusters in our sample at intermediate redshift (top panel).
(Bottom) Comparison between our results and the local measurements in DM01 (shaded regions).
The 1$\sigma$ errors on the means are plotted as solid lines (dark gray region for the local
estimates) while the scatter (rms) in each data bin is shown as a dotted line
(light gray region for local estimates).}
\label{Z_CC_vs_noCC_mean}
\end{figure}

We compute an error-weighted average $Z$ profile as done for the
temperature profile and compare it with the local measurements in DM01,
after scaling the radii to $r_{180}$ (see Figure~\ref{Z_CC_vs_noCC_mean}).
The two profiles are in agreement at $r>0.1r_{180}$, 
with both subsamples showing an evidence for a negative
gradient in metallicity significant at least at 2$\sigma$. A
different behaviour between the two subsample is observable in the
central bin, where the value in the CC sample is $\sim20\%-30\%$
higher than in the NCC sample. It is worth noticing that all the
clusters in our sample have $kT>6$ keV and this trend may be different
in lower temperature clusters.  DM01 in their analysis observed a
clear gradient in the metallicity profiles of their CC clusters, while
the profiles of their NCC clusters were almost constant. Moreover the
average metallicity observed in the CC clusters was systematically
higher than in the NCC sample, at least within $\sim0.3r/r_{180}$
from the center.  In our analysis, we do not find a clear difference
between the CC and NCC abundance profiles as in DM01. Except for the
inner radial bin where the metallicity in CC objects can be higher
by 50 per cent than in NCC ones, the CC and NCC profiles are very similar and
consistent within the errors.

To compare our mean values at intermediate redshift with the results
obtained locally from DM01, we estimate an error-weighted average
$Z$ profile for both CC and NCC objects in local and intermediate $z$ samples
(Figure~\ref{Z_CC_vs_noCC_mean}).
While the slope of the profiles is generally in agreement with DM01 for
the both the CC and NCC clusters,
the value of $Z/\langle Z \rangle$ is systematically higher in our sample
with respect to the DM01 sample, with differences up to 50\%
within $0.1 r_{180}$ of CC systems.
This might be due to the different method used to compute
$\langle Z \rangle$, being in DM01 work estimated as fit with a constant
to the radial metallicity profile.
However, apart from the inner radial region, the discrepancy between
the local profiles and the ones at intermediate redshits are
within 1$\sigma$.
Using the same method to determine $\langle Z \rangle$ on our data, we find 
that the values of $Z/\langle Z \rangle$ are fully consistent with DM01.

\subsection{Comparison with the new compilations of solar values}

All the analysis described in the current section has been performed
adopting the AG89 compilation of photospheric abundances. This choice
has been due mainly by the necessity to have a direct comparison with
previous works in the literature (e.g. DM01).  However, as explained
above in \S1, the abundance values listed in AG89 have been recently
superseeded by the new photospheric values by Grevesse \& Sauval (1998)
and A05, who introduced a 0.676 and 0.60 times lower Iron solar abundance,
respectively.  Therefore, we have also performed
the fits using solar abundances by A05 to check whether adopting
the "old" AG89 values might have introduced any bias in our analysis.\\
The shape of the $Z$ profile for both the CC sample and the NCC sample
resembles very closely that observed in Figure~\ref{Z_CC_vs_noCC_mean} for
the AG89 values of $Z$. However to better quantify this comparison we
fitted also these new values with a power-law functional.
We measure
\begin{eqnarray}
Z_{\rm CC} / \langle Z \rangle = 0.67\pm0.06\: x^{-0.25\pm0.03}; & \\ 
Z_{\rm NCC} / \langle Z \rangle = 1.00_{-0.18}^{+0.20}\: x^{-0.02\pm0.07}, &
\end{eqnarray}
with $x\equiv r/r_{180}$.
If we compare the result of the fit with the last two rows in the 
first column in Table~\ref{bestfit},
it is quite clear that the slope of the power-law in both the CC and
NCC sample is consistent with what we obtained using the AG89 values
(well within the 1$\sigma$ uncertainties) and the difference is only
in the normalization. As expected, this result might indicate that our
metallicities are mostly driven by Iron and that the contribution of
the $\alpha$ elements to the determination of the abundances is
negligible. Indeed, all our clusters have a temperature larger than 
$\sim6$ keV,
therefore the abundance measures are dominated by the Fe-K$\alpha$ line.
As a further test to this hypothesis we tried to fit the
Fe abundance independently from the $\alpha$ elements abundances.  To
this aim we used a {\tt vmekal} model where the abundances of O, Mg,
Si and S were tied-up to the same value and fitted as a single free
parameter (in order to reduce the number of free parameters), while the
other elements (apart from Fe) were frozen at solar. While the Fe
abundance remained always consistent with the value of $Z$ measured
considering the metallicity as a single parameter, in almost all the
spectra we could not find any 
statistically significant detection of a contribution from  $\alpha$
elements.  
Their abundances has been measured as upper limits
in most cases (generally $Z_\alpha<0.3$), and as very low values in
the rest of the spectra (generally $Z_\alpha\sim0.1-0.2$, often
consistent with $Z_\alpha\sim0$ at 1$\sigma$).  This is true even in
the inner part of the clusters in the sample, where the
signal-to-noise of the spectra is higher and in principle it may be
easier to detect the presence of elements
other than Iron.\\
These results suggest that the measure of $Z$ in our cluster sample
consists mainly in a measure of Iron metallicity. Therefore, adopting
the AG89 solar abundances instead of the A05 (which have different
$Z_{Fe}/Z_\alpha$ ratios) produces only a difference in the
value of the relative $Z$ measured. This does not introduce 
any bias in the analysis of the radial profiles, since 
the absolute value of $Z$ does not change and only the reference 
value assumed for the solar metallicity experience a variation.

\subsection{Gradients and cooling times}

\begin{figure}
\plotone{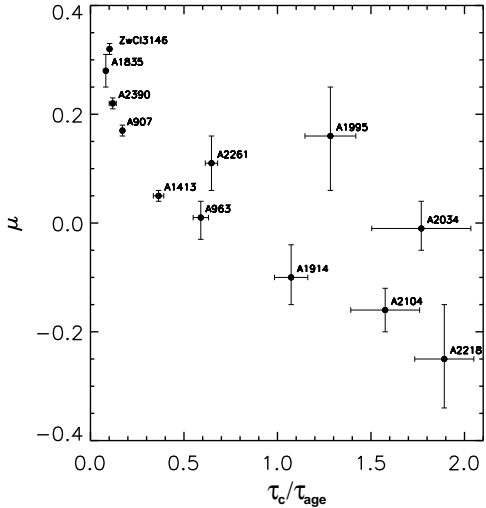}
\caption{Distribution of the exponential $\mu$ in the relation 
$kT/\langle kT \rangle=A\:(r/r_{180})^\mu$ as a function of $\tau_c/\tau_{age}$.
}
\label{alphatc}
\end{figure}

We investigate the correlation of the central slopes of the temperature 
and metallicity profiles with the central cooling time in each cluster. 
To this aim, the values of $kT(r) /\langle kT \rangle$,
$Z(r) /\langle Z \rangle$ and $r/r_{180}$ (normalized to the average
temperature, the average metallicity and the virial radius measured in
each cluster, respectively, as described in Section~3) 
with $r<0.1 r_{180}$ have been considered to characterize the cooling cores.
We find that the most robust correlation is present between 
the slope $\mu$ of the temperature profiles,
\begin{equation}
kT/\langle kT \rangle = A \:(r/r_{180})^\mu,
\end{equation}
and $\tau_c/\tau_{age}$, with a Spearman's $\rho$ rank correlation
value of -0.87 that corresponds to a significance of the
non-correlation case of $P=2\times10^{-4}$. In the $kT-Z$ and $Z-r$
relations, the values of the Spearman's $\rho$ are 0.44 and 0.54, 
corresponding to a significance of 0.15 and 0.07, respectively.
The exponent $\mu$ correlates with $\tau_c/\tau_{age}$, 
being higher at lower values of $\tau_c/\tau_{age}$,
with all the CC clusters having $\mu>0$ and $\tau_c/\tau_{age}\la0.6$
(Figure~\ref{alphatc}).

\section{Conclusions}

In the present work we analyzed a sample of 12 galaxy clusters present
in the Chandra archive with at least $\sim20,000$ net ACIS counts and
$kT>6$ keV.  These clusters were chosen in the 0.1-0.3 redshift range,
regardless of their shape. We computed the cooling time of the
clusters, subdividing the sample in 7 cool core clusters and 5
non--cool core clusters. This subdivision allowed us to compare
the two categories in a systematic fashion, following the approach of
DM01.  We performed a spectral analysis in radial bins of each cluster
in the sample requiring each bin to have $\sim7000-8000$ counts,
fitting the spectra with a thermal model with Galactic
absorption. This allowed us to derive temperature and metallicity
profiles for each cluster.  The virial radius $r_{180}$ was computed
in order to renormalize the radii to physically meaningful quantities
and investigate for self-similarities in the radial profiles. To this
aim the global temperature $\langle kT \rangle$ and metallicity
$\langle Z \rangle$ in each cluster were measured as well.
The main results coming from our work can be summarized as follows.\\
\begin{itemize}

\item The temperature profiles in the inner $0.1 r_{180}$ have, on
  average, a positive gradient, $kT(r)\propto r^\mu$ with $\mu \simeq
  0.25$ in CC systems, whereas it is almost flat in NCC systems. The
  outer regions are well fitted with a single power-law with slopes
  significantly different, being steeper ($\mu =-0.32 \pm 0.05$) in
  NCC objects.
  The general trend of our CC sample is fully consistent with 
  Vikhlinin et al. (2005) at $r\la0.3r_{180}$. The low number statistics
  above $0.3r_{180}$ prevents us from any statistically significant 
  comparison between the two samples at $r\ga0.3r_{180}$.

\item The metallicity profiles in the inner regions is almost constant
  in NCC clusters around the value measured excluding counts
  from $r<0.07 r_{180}$. In the CC sample, a steep negative gradient
  is observed ($\mu =-0.27 \pm 0.03$) in the central regions.
  At $r > 0.1 r_{180}$, a power-law reproduces well the distribution
  of the spectral measurements, with a slope that is marginally
  steeper in CC clusters ($\mu =-0.52 \pm 0.18$) than in NCC
  clusters ($\mu =-0.29 \pm 0.27$). 

\item Comparing our averaged metallicity profiles with the ones in DM01,
  we found that our values of $Z/\langle Z \rangle$ are systematically higher,
  with differences up to 50\% within $0.1 r_{180}$ of CC systems.
  This may be explained by the different method adopted in DM01 to estimate 
  $\langle Z \rangle$, as best-fit with a constant over the entire 
  metallicity profile, without any exclusion of the central core.

\item Using the solar abundances from Asplund (2005, A05) gives
  consistent results with what we obtain using the values by Anders \&
  Grevesse (1989, AG89), with a discrepancy only in the normalization 
  (as expected, $\sim60-70\%$ higher)
  but not in the slope of the $Z$ radial profiles.  Together with the
  fact that, in most cases, we were able to measure the $\alpha$
  elements only as upper limits, this indicates that our metallicities
  are mostly driven by Iron and that adopting the AG89 solar
  abundances instead of the A05 results in a difference only in the
  absolute values of the $Z$ measured but does not introduce any bias
  in the radial profile analysis.

\item Fitting a power--law shape to the temperature profiles,
  $kT/\langle kT \rangle=A\:(r/r_{180})^\mu$, we found that $\mu$
  correlates strongly with the cluster cooling times, being higher at low
  values of $\tau_c/\tau_{age}$, with all the CC clusters having
  $\alpha>0$ and $\tau_c/\tau_{age}\la0.6$. 
  As expected, strong correlation is also observed between the inner slope 
  of the metallicity profile and cluster cooling time.

\end{itemize}

In general, our results further demonstrate the invaluable role played
by X--ray archival studies of the chemo-- and thermo--dynamical
properties of galaxy clusters. Analyses based on the Chandra archive,
like that presented here (see also Vikhlinin et al. 2005; Balestra et
al. 2007; Maughan et al. 2007), in combination with analogous studies
from the XMM--Newton archive, will constitute an important heritage
from the present generation of X--ray satellites for years to
come. Nowadays, available data on the evolution of the chemical
enrichment of the ICM provide important constraints on models aimed
at explaining the past history of star formation and the dynamical
processes taking place during the cosmological build up of galaxy
clusters. However the study of the thermo-dynamical properties of the 
cooling cores and the evolution of the abundance distributions in clusters
with the redshift are just a part of what can be currently done 
exploiting in full the existing Chandra and XMM--Newton archives.
Archival works like ours have the potential to shed new light on the 
properties of the
stellar populations responsible for the ICM enrichment, and on the
mechanisms which lead to the generation of the cool cores and
determine the transport and diffusion of heavy elements from star
forming regions. 

\acknowledgments AB and PM acknowledge financial support from CXO
grant AR6-7015X and from NASA grant GO5-6124X. 
We acknowledge financial contribution from contract
ASI--INAF I/023/05/0. PT and SB acknowledge financial
contribution from the PD51 INFN grant.
We thank A. Vikhlinin for providing us the temperature and abundance 
profiles of some of the clusters in his sample.
We also thank F.Gastaldello for useful discussions.
We thank the anonymous referee for comments and suggestions useful 
to improve the presentation of the paper.

\end{document}